# Longitudinal optical force of laser pulses in continuous media

L. Kovachev*

Institute of Electronics, Bulgarian Academy of Sciences, 72 Tzarigradsko choose ,1784 Sofia, Bulgaria

**ABSTRACT**

From long time transverse and longitudinal optical forces are used for non-contact and noninvasive manipulation of small particle. The following question arises: What is the impact of these forces on the continuous media as air and silica. In this work we obtain analytical expressions for radiation force and radiation potential in dipole approximation of laser pulses propagating in dielectric media. The longitudinal force is proportional to the second time derivative of the intensity profile. At fixed initial energy of a laser pulse, as shorter is the pulse as bigger is the force. It is possible the neutral particles in gases and solids to be confined in the pulse envelope and to move with group velocity. In silica the longitudinal force of a femtosecond pulse is of few order of magnitude greater than molecular forces. Thus, the ablation in silica can be realized through broken molecular connections.

**Keywords:** Lasers; Lasers trapping, Ablation,

## 1. INTRODUCTION

As was demonstrated by Ashkin[1,2], it is possible to trap particles by lasers, working in CW regime. The analytical expression of the radiation force is obtained in dipole approximation and, as is well known, is proportional to the transverse gradient of the square of electrical field. Recently, were obtained results of manipulating Rayleigh dielectric particles by optical pulses[3-8]. The theoretical and experimental results in these investigations show that in pulse regime additional longitudinal force exists. This longitudinal force can be greatly enhanced when the pulse duration decreases. The theoretical results in these investigations do not use the fact that the optical response of a laser pulse is non-stationary in dielectric media. For this reason, the radiation force is presented by the phase velocity of the pulse. The flow of energy and the pulse envelope of one pulse propagate with group velocity. This fact must be taking into account in the calculation of the ponder-motor force. The question what kinds of radiation forces exist for laser pulses propagate in continuous dielectric media is still open. In this paper we obtain analytical expression for the longitudinal radiation force of laser pulse propagating in dielectric media. This force is proportional to the second derivative of pulse time envelope. That is why the force vanishes in CW regime, while in the femtosecond region leads to trapping of particles in dielectric media into the pulse envelope. The moving particles admit an unexpected nonlinear response and the result is new nonlinear evolution of the laser pulses. Our calculations show that even with nJ femtosecond pulse propagating in silica, the longitudinal force is of few order of magnitude greater than molecular forces.

*lubomirkovach@yahoo.com

## 2. LONGITUDINAL RADIATION FORCE DENSITY IN AMPLITUDE APPROXIMATION

In dipole approximation, as it was shown by Gordon[9] the radiation force is simply Lorentz force and can be rewritten in the form

$$\vec{F} = \alpha \left[ \frac{1}{2} \nabla \left( \vec{E}^2 \right) + \frac{d}{dt} \left( \vec{E} \times \vec{B} \right) \right] = \alpha \left[ \frac{1}{2} \nabla \left( \vec{E}^2 \right) + \frac{4\pi}{c} \frac{d}{dt} \vec{S} \right], \tag{1}$$

where $\alpha$ is the atomic polarizability and $\vec{S} = \vec{E} \times \vec{B}$ is the Pointing vector. The first term in the brackets is the well known gradient force, while the second is associated with the propagation of the pulse energy and is proportional to the Pointing vector. To obtain ponder-motor force density in dielectrics we multiply $\alpha$ by the number of atoms per volume $N$ and additionally use the local field correction

$$\chi^{(1)} = \frac{N\alpha}{1 - \frac{4}{3}\pi N\alpha}. \tag{2}$$

The force per unit volume becomes

$$<\vec{F}> = \chi^{(1)}\left[\frac{1}{2}\nabla(\vec{E}^2) + \frac{4\pi}{c}\frac{d}{dt}\vec{S}\right]. \tag{3}$$

One natural way to include the media parameters into the expression of Ponder-Motor (PM) force density (3) is by using the divergence of the Pointing vector.

$$-\frac{c}{4\pi}\nabla\cdot\vec{S} = \frac{1}{4\pi}\left(\vec{E}\cdot\frac{\partial}{\partial t}\vec{D} + \vec{H}\cdot\frac{\partial}{\partial t}\vec{B}\right), \tag{4}$$

where for dielectrics with non-stationary linear response it is fulfilled

$$\vec{D} = \vec{E} + \vec{P}_{lin}; \qquad \vec{P}_{lin} = 4\pi\int_0^\infty R^{(1)}(\tau)\vec{E}(t-\tau,r)d\tau, \tag{5}$$

$$\vec{H} = \vec{B}. \tag{6}$$

Lets us present the electrical and magnetic fields of the laser pulse by the pulse envelopes and carrying frequency $\omega_0$

$$\vec{E} = (\vec{A}(x,y,z,t)\exp(i\omega_0 t) + c.c.)/2, \tag{7}$$

$$\vec{H} = \vec{B} = (\vec{C}(x,y,z,t)\exp(i\omega_0 t) - c.c)/2i, \tag{8}$$

where $\vec{A}(x,y,z,t)$ and $\vec{C}(x,y,z,t)$ are the complex amplitudes of electrical and magnetic fields correspondingly. After using the Fourier presentation of equations (4) - (8), and developing the product of dielectric constant with frequency $\omega\varepsilon(\omega)$ in Tailor series near $\omega_0$.

$$\omega\varepsilon(\omega) = \omega_0\varepsilon(\omega_0) + \frac{\partial[\omega\varepsilon(\omega)]}{\partial\omega}\bigg|_{\omega=\omega_0}(\omega-\omega_0), \tag{9}$$

and after short calculations we obtain the following expression for the divergence of the Pointing vector

$$\nabla\cdot\vec{S} = -\frac{1}{4v_{gr}}\frac{\partial|\vec{A}(r,t)|^2}{\partial t} - \frac{\mu(\omega_0)}{4c}\frac{\partial|\vec{C}(r,t)|^2}{\partial t}, \tag{10}$$

where $c$ is the light velocity in vacuum, $v_{gr} = c/\left(\varepsilon(\omega_0) + \omega_0\frac{\partial[\varepsilon(\omega)]}{\partial\omega}\bigg|_{\omega=\omega_0}\right)$ is the group velocity and $\mu(\omega_0)$ is the magnetic permeability. In dipole approximation we use the first term of the right-hand side of equation (10) associated with the amplitude $\vec{A}(x,y,z,t)$ of electrical field. It is not difficult to show that in paraxial approximation, after using the first Maxwell equation, the second term from the right-hand side associated with the magnetic field gives the same amount of (asymmetric) energy flow as the first one and propagates also with group velocity.

The next step is to use the differences between the atom and optical scales. The atoms and molecules can be characterized by their atom (molecular) response of order of $\tau_0 \approx 2-3$ fs. During this time the laser pulse propagates at distance of $z_{resp} = \tau_0 v_{gr} \approx 0.5-1.$ μm. The optical scale is characterized by diffraction length and for a typical laser pulse varies $z_{diff} = k_0 d_0^2 \approx 15-150$ cm. Since $z_{diff} >> z_{resp}$ always at one diffraction length there are thousand oscillations of the atom dipole. Thus, the shape of the pulse does not change significantly at distances less than one diffraction length and for these distances we can use the following approximation

$$\vec{S} = [0,0,S_z]; \qquad \frac{\partial S_z}{\partial z} = -\frac{1}{4v_{gr}} \frac{\partial |\vec{A}(x,y,z-v_{gr}t)|^2}{\partial t} \qquad (11)$$

where $z$ is the direction of propagation of the laser pulse. The equation (11) presents actually the flow of energy trough a plane surface situated at the point $z = 0$ and orthogonal to the direction of laser pulse propagation. The coordinates of the intensity of the pulse form the left hand of this surface are $z - v_{gr}t; \quad t > 0$, while from the right-hand side are $z + v_{gr}t; \quad t > 0$. Integrating the equation (11) we obtain following expression for the Pointing vector

$$S_z = -\frac{1}{4v_{gr}} \int_{-\infty}^{0} \frac{\partial |\vec{A}(x,y,z-v_{gr}t)|^2}{\partial t} dz - \frac{1}{4v_{gr}} \int_{0}^{\infty} \frac{\partial |\vec{A}(x,y,(z+v_{gr}t))|^2}{\partial t} dz =$$
$$-\frac{1}{2v_{gr}} \int_{-\infty}^{0} \frac{\partial |\vec{A}(x,y,z-v_{gr}t)|^2}{\partial t} dz. \qquad (12)$$

Thus, from equations (3) and (12) we obtain for the longitudinal part of the ponder-motor force density

$$<F_z> = \frac{4\pi\chi^{(1)}}{c} \frac{\partial S_z}{\partial t} = -\frac{2\pi\chi^{(1)}}{cv_{gr}} \int_{-\infty}^{0} \frac{\partial^2 |\vec{A}(x,y,z-v_{gr}t)|^2}{\partial t^2} dz. \qquad (13)$$

In equation (13) the longitudinal part of force density depends on the second derivative of intensity profile of the laser pulse. In CW regime this derivative vanishes and there is only transverse gradient forces investigated in[1-2]. As shorter is the pulse as bigger is the longitudinal force. In in femtosecond region, as we will see below, this force takes significant values.

## 3. LONGITUDINAL RADIATION FORCE DENSITY AND POTENTIAL IN APPROXIMATION OF FIRST ORDER OF DISPERSION

The difference between the atomic and optical scales gives the opportunity at decade's centimeters, where the shape of the pulse is maintained the integral of force density (13) to be solved. In first approximation of dispersion one initial Gaussian pulse propagates with group velocity and preserves its shape

$$A(t,x,y,z) = A_0 \exp\left(-\frac{x^2 + y^2}{2d_0^2} - \frac{(z-v_{gr}t)^2}{2z_0^2}\right). \qquad (14)$$

After substituting Eq. (14) in Eq. (13), differentiating twice by time and integrating by the variable $z$ we obtain

$$<F_z> = -\frac{4\pi\chi^{(1)}}{c}\frac{A_0^2}{t_0}\exp\left(-\frac{x^2+y^2}{d_0^2}\right)\exp\left(-\frac{t^2}{t_0^2}\right)\frac{t}{t_0}. \qquad (15)$$

The paraxial optics works in spatio-temporal coordinates. That is why the expression for PM density force is in the same coordinates. There are two possibilities to present PM force in a real coordinate system $(x, y, z)$. The first one is to solve paraxial spatio-temporal equation in Galilean frame. The second is to use the relations $z = v_{gr}t$ and $z_0 = v_{gr}t_0$. In this way we obtain the real 3D shape of the radiation force. In addition we present the squared modulus of electrical field by the intensity

$$|A_0|^2 = \frac{2\pi}{cn(\omega_0)}I_0. \qquad (16)$$

The expression for the PM density force in a real 3D space is transformed to

$$<F_z> = -\frac{8\pi^2\chi^{(1)}v_{gr}}{n(\omega_0)c^2}\frac{I_0}{z_0}\exp\left(-\frac{x^2+y^2}{d_0^2}\right)\exp\left(-\frac{z^2}{z_0^2}\right)\frac{z}{z_0}. \qquad (17)$$

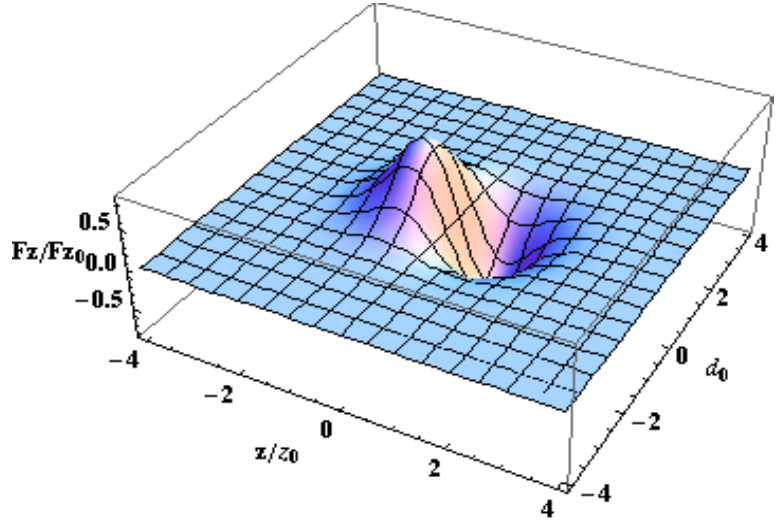

Fig.1 Graphics of the PM longitudinal force density of a single laser pulse. The pulse front attracts the ensemble of particles to the center of the pulse while the back side pushes them again to the center.

The longitudinal PM force propagates with group velocity. The 3D image of the PM force density is plotted on Fig. 1. The pulse front attracts the ensemble of particles to the center of the pulse while the back side pushes them again to the center. The $F_z$ force depends on 3D coordinates and a potential density can be introduced by

$$U(x,y,z) = \int_{-\infty}^{z} F_z dz. \qquad (18)$$

The result is

$$U_z = \int_{-\infty}^{z} F_z dz = -\frac{4\pi^2\chi^{(1)}v_{gr}I_0}{n(\omega_0)c^2}\exp\left(-\frac{x^2+y^2}{d_0^2}\right)\exp\left(-\frac{z^2}{z_0^2}\right); \quad \left[\frac{dynes}{cm^3}\right]. \qquad (19)$$

Graph of the potential density is plotted on Fig 2. The Gaussian shape of the pulse plays the role of an attractive potential.

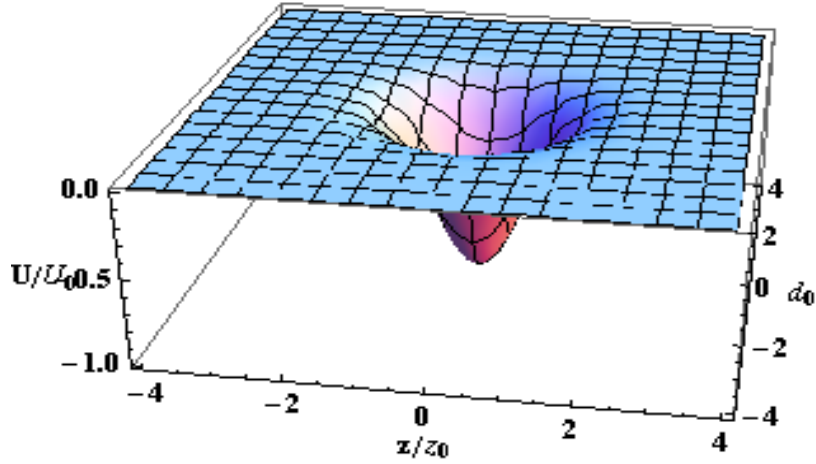

Fig 2. Graphics of the potential density of a Gaussian laser pulse. The shape of the pulse, moving with the group velocity plays the role of an attractive potential.

## 4. VALUES OF THE LONGITUDINAL RADIATION FORCE AND POTENTIAL

In the previous section we obtained formulas for PM longitudinal radiation force density and potential in approximation of first order of dispersion for a single Gaussian laser pulse. The formulas for the force density (17) and the potential density (18) for Gaussian pulses can be integrated over the whole space. As a result we will obtain after integrating the normalized Gaussian pulses only two additional constants - the spot of the pulse $d_0$ and the longitudinal shape $z_0 = v_{gr} t_0$. In this way we obtain an effective real force and a potential at level of the spot diameter $d_0$ and in the frame of its longitudinal shape $z_0$.

For the longitudinal radiation force after integrating (17) from $z = -\infty$ to $z = 0$ (the back side of the pulse will give the same amount of force but with opposite direction) we have

$$F_z^{eff} = \iiint_{x,y,z} F_z dxdydz = -\pi \frac{\chi^{(1)}}{n(\omega_0)} \frac{v_{gr}}{c^2} \frac{E_0^{laser}}{t_0} = -Const \frac{E_0^{laser}}{t_0}, \quad (20)$$

where with $E_0^{laser}$ is signed the energy of the initial laser pulse. The expressions (20) and all formulas below are written in MKS units ($\chi_{MKS}^{(1)} = 4\pi \chi_{gaussian}^{(1)}$). It is important to mention here the dependence of the longitudinal PM force on the pulse time duration. As the pulse is getting shorter the force becomes stronger.

The expression of the potential after integration becomes

$$U_z^{eff} = F_z^{eff} z_0 = -\pi \frac{\chi^{(1)}}{n(\omega_0)} \frac{v_{gr}^2}{c^2} E_0^{laser}. \quad (21)$$

How deep is the radiation potential in air? Let's compare it to the Boltzmann energy of free particles at room temperature $T = 300\ K$. The value is

$$U_B = k_B T = 4.14 \times 10^{-21}\ [J]. \quad (22)$$

In our example we use laser pulse having initial energy in range $E_0^{laser} \cong 1 \mu J$. The potential is

$$U_z^{eff} \cong 1. \times 10^{-9} \quad [J], \tag{23}$$

which is twelve orders of magnitude greater than the Boltzmann energy. The Boltzmann factor is very small

$$R = \exp\left(-\frac{U_{max}}{k_B T}\right) << 1. \tag{24}$$

This results show that self-confinement of particles in the pulse envelope is possible. Let us suppose that in isotropic media the particles are really confined in the pulse envelope. Then a quite interesting nonlinear response can be observed. The captured particles will generate not at third harmonics $3\omega_0 = 3k_0 v_{ph}$, but at frequency proportional to the three times group-phase velocity difference $3\omega_{THz} = 3k_0(v_{ph} - v_{gr})$. Such a generation indeed was observed in recent experiments[10].

How strong is the force in fused silica for example? The typical molecular forces in silica are of order of

$$F_{mol}^{silica} \approx -10 \quad \left[\frac{eV}{A°}\right] \cong -1.6 \times 10^{-8} \quad [N]. \tag{25}$$

If we use a laser pulse having time duration $t_0 = 100 \quad [fs]$ and energy $E_0^{laser} \cong 100 \quad [nJ]$ the value of the longitudinal force becomes

$$F_{pm}^{pulse} \cong -2.6 \times 10^{-3} \quad [N]. \tag{26}$$

This value is five order of magnitudes greater than the molecular forces in silica. This calculations show that the ablation obtained by femtosecond pulses in silica can be result of broken molecular connections due to longitudinal radiation force.

## 5. CONCLUSIONS

Up to now, the basic experimental and theoretical investigations are related to the study of radiation forces produced by laser beams and pulses acting on individual Rayleigh dielectric particles. In this paper we explore the impact of this force on continuous media and dielectrics. Thus, the individual force applied to an atom is transformed to density force per volume. The optical response of dielectric media connected with the propagation of laser pulses is non-stationary and also is taken into account. As result, we obtain analytical expression for the longitudinal density force and density potential of Gaussian laser pulse propagating in dielectric media. It is possible to integrate these densities using the paraxial approximation in optics. As result we obtain effective longitudinal potential and force, acting in the media at level of the pulse spot. The longitudinal radiation force depends on the time duration of the laser pulse. As the pulse is getting shorter the force becomes stronger. In the femtosecond region this force in silica is of few orders of magnitude greater than the molecular forces. Therefore, the ablation in silica can be realized by broken molecular connections due to the PM force. In air the longitudinal potential of Gaussian laser pulse with energy $E_0^{pulse} \cong 1.$ μJ is of twelve orders of magnitude greater than the Boltzmann energy of free particles. It is possible that neutral particles to be confined in the pulse envelope and to move with group velocity. In nonlinear regime of propagation this leads to new nonlinear conversion mechanisms.

## 6. ACKNOWLEDGMENTS

This material is based upon work supported by the Air Force Office of Scientific Research under award number FA9550-19-1-7003. The present work is funded also by Bulgarian National Science Fund by grant DN18/11.